 \documentclass[preprint,review,12pt]{elsarticle}

\usepackage{amssymb}
\usepackage{graphicx}
\usepackage{bm}

\newcommand{\bd}{\begin{displaymath}}
\newcommand{\ed}{\end{displaymath}}
\newcommand{\be}{\begin{equation}}
\newcommand{\ee}{\end{equation}}
\newcommand{\beq}{\begin{eqnarray}}
\newcommand{\eeq}{\end{eqnarray}}  
\newcommand{\beqs}{\begin{eqnarray*}}
\newcommand{\eeqs}{\end{eqnarray*}}  

\journal{Journal of Magnetism and Magnetic Materials}

\begin{document}

\begin{frontmatter}



\title{Magnetocaloric effect in \{[Fe(pyrazole)$_4$]$_2$[Nb(CN)$_8$]$\cdot$4H$_2$O\}$_n$ molecular magnet}


\author[ad1]{R. Pe\l{}ka\corref{cor1}}
\cortext[cor1]{Corresponding author, e-mail: robert.pelka@ifj.edu.pl}
\author[ad1]{P. Konieczny}
\author[ad1]{P. M. Zieli\'{n}ski}
\author[ad1]{T. Wasiuty\'{n}ski}
\author[ad2]{Y. Miyazaki}
\author[ad2]{A. Inaba}
\author[ad3]{D. Pinkowicz}
\author[ad3]{B. Sieklucka}
\address[ad1]{The H. Niewodnicza\'{n}ski Institute of Nuclear Physics Polish Academy of Sciences, Radzikowskiego 152, 31-342 Krak\'{o}w, Poland}
\address[ad2]{Research Center for Structural Thermodynamics, Graduate School of Science, Osaka University, Toyonaka, Osaka 560-0043, Japan}
\address[ad3]{Faculty of Chemistry, Jagiellonian University, Ingardena 3, 30-060 Krak\'{o}w, Poland}

\begin{abstract}
Magnetocaloric effect in \{[Fe(pyrazole)$_4$]$_2$[Nb(CN)$_8$]$\cdot$4H$_2$O\}$_n$  molecular magnet is reported. It crystallizes in tetragonal I4$_1$/a space group. The compound exhibits a phase transition to a long range magnetically ordered state at $T_\mathrm{c}\approx$8.3\, K. The magnetic entropy change $\Delta S_\mathrm{M}$ as well as the adiabatic temperature change $\Delta T_\mathrm{ad}$ due to applied field  change $\mu_0\Delta H$=0.1, 0.2, 0.5, 1, 2, 5, 9\,T as a function of temperature have been determined by the relaxation calorimetry measurements. The maximum value of $\Delta S_\mathrm{M}$ for $\mu_0\Delta H=5$\,T is 4.9 J mol$^{-1}$ K$^{-1}$ (4.8 J kg$^{-1}$ K$^{-1}$) at 10.3\, K. The corresponding maximum value of $\Delta T_\mathrm{ad}$ is 2.0\,K at 8.9\,K. The temperature dependence of the exponent $n$ characterizing the field dependence of $\Delta S_\mathrm{M}$ has been estimated. It attains the value of 0.64 at the transition temperature, which is consistent with the 3D Heisenberg universality class.  
\end{abstract}


\end{frontmatter}



\section{\label{section1} Introduction}
Magnetocaloric effect (MCE), i.e. the thermal change of a magnetic material when exposed to applied magnetic field, is an issue of intensive research not only due to the fact that magnetic refrigerators have an enhanced efficiency with respect to conventional gas compression-expansion ones but they are also more environmentally friendly, as they do not require gasses associated with greenhouse effect or ozone depletion \cite{zimm}. Current developments in materials science related to this field involve on the one hand the enhancement of materials performance mostly realized through giant magnetocaloric effect (GMCE) \cite{gschneidner, bruck} and the cost reduction through replacement of rare earth elements by transition metal alloys \cite{tegus}. However, although GMCE shown by materials exhibiting first-order magnetostructural phase transitions delivers a large magnetic entropy change $\Delta S_\mathrm{M}$ it has two important drawbacks, namely, the narrowness of the $\Delta S_\mathrm{M}$ vs $T$ curve and the presence of hysteresis leading to low operational frequencies and limited cooling power (refrigerant capacity RC). Compounds undergoing second-order phase transitions do not show thermal hysteresis and their $\Delta S_\mathrm{M}(T)$ is extended in a wider temperature range, albeit it is smaller in magnitude. Therefore, a compromise between an optimal RC and the lack of thermal hysteresis makes them better candidates for the development of magnetic cooling devices at the present moment. 

The investigation of MCE is also important from the fundamental viewpoint, as the dependence of $\Delta S_\mathrm{M}$ on temperature and magnetic field change $\Delta H$ represents a unique characterstics of a material correlated to its critical behaviour. In the present paper we report a study of MCE in a bimetallic molecular magnet  \{[Fe$^\mathrm{II}$(pyrazole)$_4$]$_2$[Nb$^\mathrm{IV}$(CN)$_8$]$\cdot$4H$_2$O\}$_n$ (pyrazole is a five membered C$_3$H$_4$N$_2$ ring ligand). MCE in molecular magnets has been investigated most of all for single molecule magnets (SMMs), where a substantial entropic effect was anticipated due to their large ground-state spin value \cite{manoli1,manoli2,evangelisti1,evangelisti2}. Besides, molecular rings, forming a subclass of molecular magnets characterized by a typical cyclic shape and a dominant antiferromagnetic coupling between nearest neighbouring ions, provide the possibility for obtaining collections of (super-)paramagnetic particles with suitable thermal properties \cite{affronte}.  Apart from that, first studies of MCE driven by the transition to a long-range magnetically ordered phase dealt with Prussian blue analogues \cite{manuel, sharma, yusuf}. Recent examples refer to an interesting instance of a molecular sponge changing reversibly the ordering temperature and the coersive field upon hydration/dehydration \cite{fitta1}, and a couple of molecular magnets, bimetallic octacyanoniobates \{[M$^\mathrm{II}$(pyrazole)$_4$]$_2$[Nb$^\mathrm{IV}$(CN)$_8$]$\cdot$4H$_2$O\}$_n$ (M=Mn, Ni) isomorphous with the compound under study \cite{fitta2}.

\{[Fe$^\mathrm{II}$(pyrazole)$_4$]$_2$[Nb$^\mathrm{IV}$(CN)$_8$]$\cdot$4H$_2$O\}$_n$ crystallizes in the tetragonal space group I4$_1$/a \cite{pinkowicz}. Its unique structure consists of a 3D skeleton, where each Nb$^\mathrm{IV}$ centre is linked through the cyanido bridges Fe$^\mathrm{II}$-NC-Nb$^\mathrm{IV}$ to four Fe$^\mathrm{II}$ ions, whereas each Fe$^\mathrm{II}$ centre is bridged to only two Nb$^\mathrm{IV}$ ions. The remaining part of the six-membered coordination sphere of Fe$^\mathrm{II}$ is filled with pyrazole molecules, while the Nb$^\mathrm{IV}$ ion coordinates further four terminal CN$^{-}$ ligands. It is a unique structural feature that such low connectivity indices produce a 3D extended network. The graphical representation of the crystal structure of the compound may be found in \cite{konieczny}. The compound undergoes a magnetic transition to a long-range order state at $T_\mathrm{c}\approx 7.8$\,K (inferred from magnetic data) \cite{konieczny}. Recently, a strong evidence for the assignment of the compound to the universality class of the 3D Heisenberg model has been provided \cite{konieczny}. The analysis of the dc susceptibility and isothermal magnetization carried out in the framework of the mean-field model suggested that the character of the exchange coupling between the Fe$^\mathrm{II}$ and Nb$^\mathrm{IV}$ centres is antiferromagnetic \cite{pinkowicz}.  

The spin of the Fe$^\mathrm{II}$ ion is $S_\mathrm{Fe}=2$, while the spin carried by the Nb$^\mathrm{IV}$ ion is $S_\mathrm{Nb}=1/2$. Hence the maximal molar magnetic entropy of the system amounts to $S_\mathrm{M}^\mathrm{max}=R\ln\left(2S_\mathrm{Fe}+1\right)^2\left(2S_\mathrm{Nb}+1\right)\approx 32.53$ J K$^{-1}$ mol$^{-1}$.  This preliminary calculation and the fact that the compound exhibits a second-order phase transition makes us expect a considerable magnetocaloric response. On the other hand, the preliminary analysis of the magnetometric data pointed to an antiferromagnetic coupling between the Fe$^\mathrm{II}$ and Nb$^\mathrm{IV}$ ions, which makes the total spin of $S=2S_\mathrm{Fe}-S_\mathrm{Nb}=7/2$ a more appropriate variable at least in the magnetic fields below the decoupling threshold. The entropy content associated with this spin value amounts to $R\ln(2S+1)\approx 17.29$ J K$^{-1}$ mol$^{-1}$, which is reduced by half but still substantial.

\section{\label{section2}Results}
The heat capacity measurements were carried out with the PPMS Quantum Design instrument by the relaxation calorimetry technique. The sample of mass 3.3189\,mg was pressed to form a small pellet. The measurements were performed in the cooling direction in the temperature range 20.2-0.36\,K in the applied field of $\mu_0 H=$0.1, 0.2, 0.5, 1, 2, 5, and 9\,T. The two main MCE characteristics of the compound, i.e. the temperature dependences of the isothermal entropy change $\Delta S_\mathrm{M}$ and the adiabatic temperature change $\Delta T_\mathrm{ad}$, were determined directly using the measured heat capacity values.  $\Delta S_\mathrm{M}$ was calculated using the formula
\be
\Delta S_\mathrm{M}=\int_0^T\frac{C_p(T^\prime,H=0)-C_p(T^\prime,H)}{T^\prime}\mathrm{d}T^\prime,
\label{eq1}
\ee
which obviates the need to determine the normal heat capacity. However, it does require the extrapolation of the $C_p$ data down to the zero temperature. The extrapolation was performed assuming a two parameter function $C_{p,\mathrm{LT}}=AT^3+BT^{3/2}$, where the first term corresponds to the lattice contribution, while the second term represents the Bloch law for the magnon contribution. The adiabatic temperature change $\Delta T_\mathrm{ad}$ was estimated on the basis of the formula
\be
\Delta T_\mathrm{ad}=\left[T(S,H)-T(S,H=0)\right]_S
\label{eq2}
\ee 
which requires the inversion of the $S(T)$ dependence. The temperature dependence of the entropy was estimated applying the same extrapolation scheme. 
\begin{figure}[!ht]
\centering
\includegraphics[width=\textwidth]{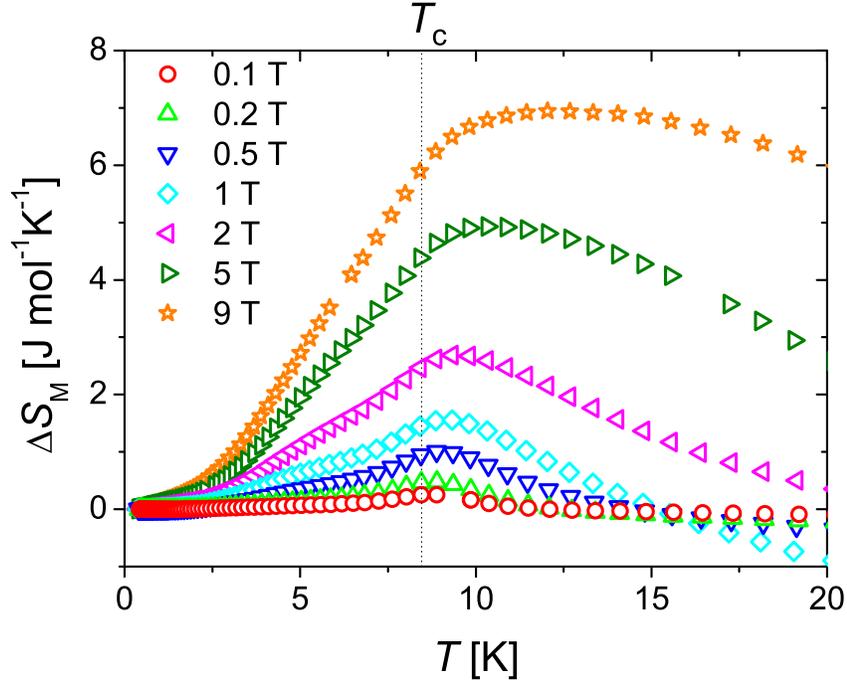}
\caption{Temperature dependence of the entropy change $\Delta S_\mathrm{M}$. }
\label{fig1}
\end{figure}

Figure \ref{fig1} shows $\Delta S_\mathrm{M}$ as a function of temperature for an array of magnetic field changes $\mu_0\Delta H=$0.1, 0.2, 0.5, 1, 2, 5, and 9\,T. Table \ref{tab1} lists the values of $\Delta S_\mathrm{M}$ for $\mu_0\Delta H=$2, 5, and 9 \,T. The value of $\Delta S_{\mathrm{M}}^{\mathrm{peak}}$=4.9 J mol$^{-1}$ K$^{-1}$ detected for $\mu_0\Delta H=5$\,T is lower than those observed for the same field change for the isostructural compounds  \{[M$^\mathrm{II}$(pyrazole)$_4$]$_2$[Nb$^\mathrm{IV}$(CN)$_8$]$\cdot$4H$_2$O\}$_n$  with M=Ni (6.1 J mol$^{-1}$ K$^{-1}$) and M=Mn (6.7 J mol$^{-1}$ K$^{-1}$) \cite{fitta2}. The relatively high value for M=Mn is probably related to the high spin value of 5/2 of the Mn$^\mathrm{II}$ ion. Although the spin value of the Ni$^\mathrm{II}$ centre ($S_\mathrm{Ni}=1$) is lower than that of the Fe$^\mathrm{II}$ ion ($S_\mathrm{Fe}=2$) the compound with M=Ni shows a larger MCE effect. This may be due to both the presumed anisotropy of the Fe$^\mathrm{II}$ centre which is known to affect adversely the MCE effect \cite{evangelisti1} as well as the fact that the exchange coupling in the compound containing  Ni$^\mathrm{II}$ is of ferromagnetic character \cite{pinkowicz}. 

Above about 15\,K for the low values of the magnetic field change the sign of $\Delta S_\mathrm{M}$ is negative. The amplitude of this reverse MCE increases with increasing magnetic field. The deepest downshift of the MCE curve is observed for the magnetic field change of 1\,T. However, for the following field change of 2\,T this effect vanishes as it does for still higher field values.  We nevertheless cannot unambiguously point to the reason of that effect. We have never encountered a similar behaviour in the literature. We suppose that it might be attributed to the presumed local anisotropy of the Fe(II) ion. In particular, for an easy-plane anisotropy the entropy of the spin system in zero applied field may be at sufficiently low temperatures lower than that in a nonvanishing external field oriented along the anisotropy axis which is equivalent to the reverse MCE. Unfortunately, not possessing the single crystal sample, we are unable to judge if this anisotropy effect would be strong enough to appear at the temperatures as high as 15 K and above.

\begin{figure}[!ht]
\centering
\includegraphics[width=\textwidth]{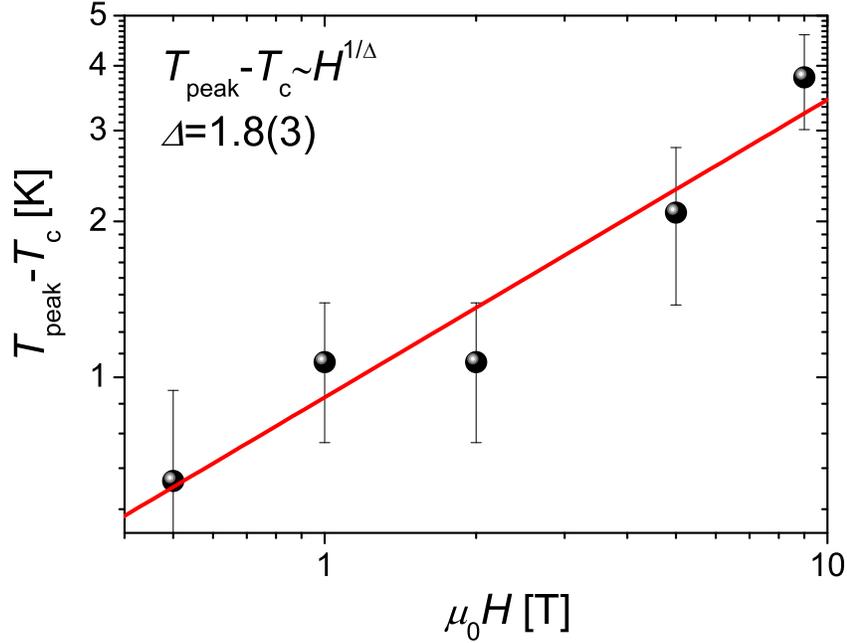}
\caption{Log-log plot of the field dependence of the distance $T_\mathrm{peak}-T_\mathrm{c}$. }
\label{fig2}
\end{figure}

From Fig. \ref{fig1} it is apparent that the peak values shift off from the critical temperature $T_\mathrm{c}=8.3$\,K (inferred from the position of the heat capacity anomaly) to higher temperatures with increasing magnetic field change. This shift can be understood from an experimental point of view, where the determination of the Curie temperature from magnetization curves usually implies the determination of their inflection points, which shift to higher temperatures for increasing magnetic fields and give rise to an apparent increase in $T_\mathrm{c}$ for higher magnetic fields. The distance  between $T_\mathrm{c}$ and $T_\mathrm{peak}$ increases with field  following a power law $H^{1/\Delta}$ (with the so called gap exponent $\Delta$), as predicted by scaling laws \cite{franco1}.
\begin{table}
\begin{center}
    \begin{tabular}{c|c|c}                   
     & $\Delta S_{\mathrm{M}}^{\mathrm{peak}}$ [J K$^{-1}$ mol$^{-1}$]  & $\Delta T_{\mathrm{ad}}^{\mathrm{peak}}$ [K] \\
     & ($T_\mathrm{peak}$ [K]) & ($T_\mathrm{peak}$ [K])
     \\ \hline
        $\mu_0\Delta H=2$\,T  & 2.7 (9.3)    & 1.1 (8.8)   \\ 
        $\mu_0\Delta H=5$\,T  & 4.9 (10.3)  & 2.0 (8.9)   \\ 
        $\mu_0\Delta H=9$\,T  & 6.9 (12.1)  & 2.8 (8.8)    \\
    \end{tabular}
    \caption{Peak values of $\Delta S_\mathrm{M}$ and $\Delta T_\mathrm{ad}$  for $\mu_0\Delta H=$2, 5, and 9\,T.}    \label{tab1}
    \end{center}
\end{table}
This feature provides insight into the critical behaviour of the compound. Figure \ref{fig2} shows the log-log plot of the field dependence of the distance $T_\mathrm{peak}-T_\mathrm{c}$. The lowest field data have been omitted because of the large uncertainty of the peak position. The slope of the linear function fitted to the data yields the inverse of the gap exponent which implies $\Delta=1.8(3)$. 

\begin{figure}[!ht]
\centering
\includegraphics[width=\textwidth]{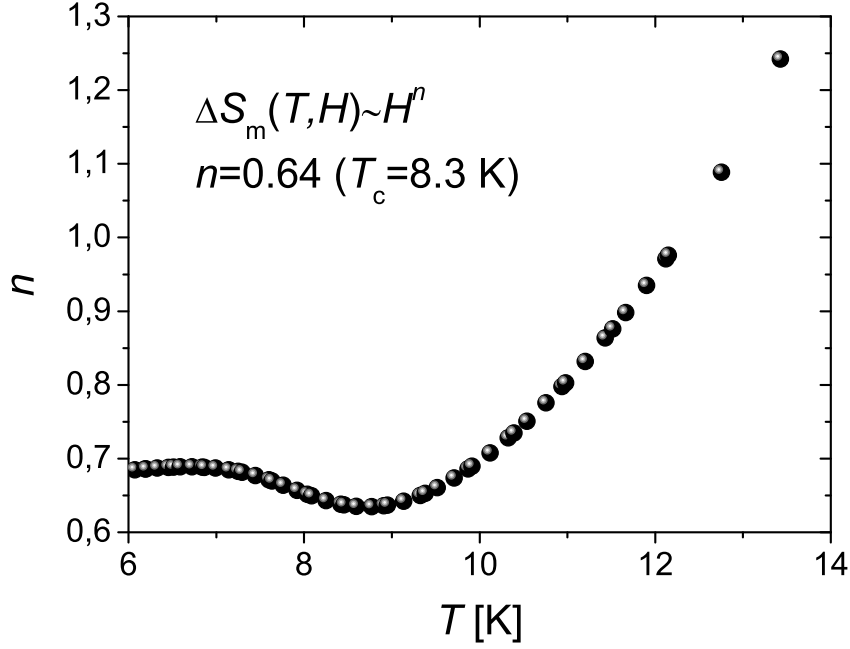}
\caption{Temperature dependence of exponent $n$. }
\label{fig3}
\end{figure}

\begin{figure}[!ht]
\centering
\includegraphics[width=\textwidth]{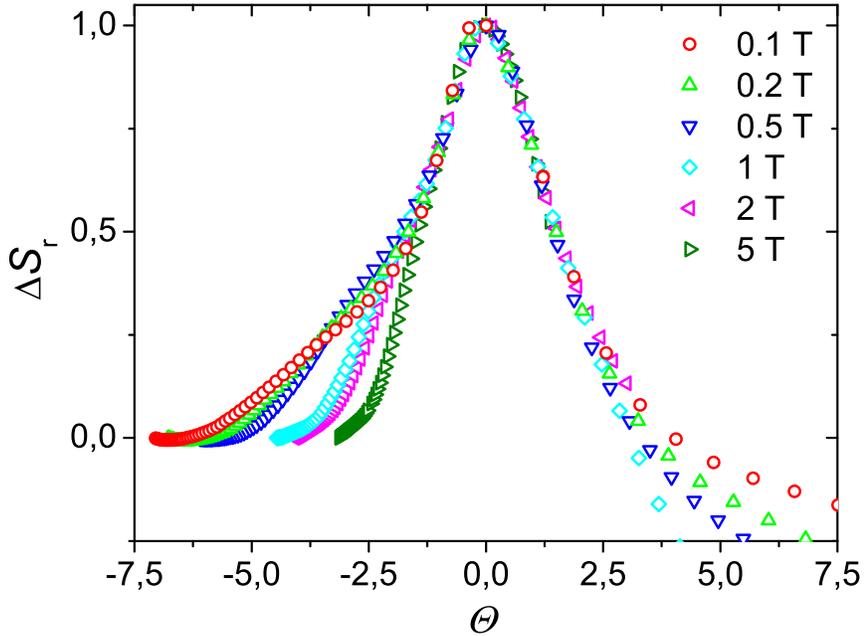}
\caption{An attempt to determine a universal curve of the entropy change $\Delta S_\mathrm{M}$ of the compound.}
\label{fig4}
\end{figure}

\begin{figure}[!ht]
\centering
\includegraphics[width=\textwidth]{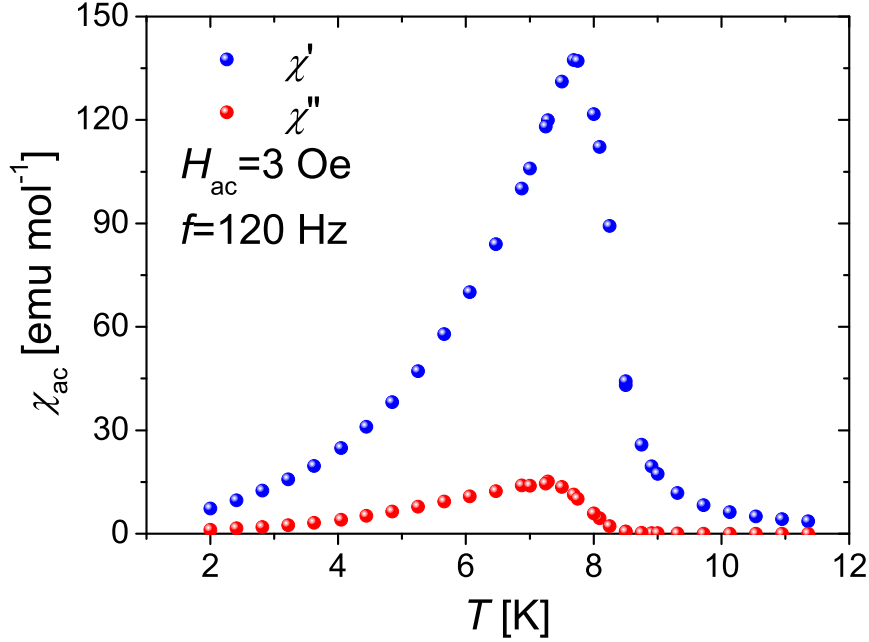}
\caption{Temperature dependence of ac susceptibility.}
\label{fig5}
\end{figure}

The field dependence of MCE has been studied intensively either experimentally \cite{casanova, tishin} or from a theoretical viewpoint by using the description within the mean-field model \cite{oesterreicher} or applying the equation of state for materials with a second-order magnetic phase transition \cite{franco1,franco2}. Expressing the field dependence of the entropy change as the power law $\Delta S_\mathrm{M}\sim H^n$, a relationship between the exponent $n$ at the Curie temperature and the critical exponents of the material has been established
\be
n|_{T=T_\mathrm{c}}=1+\frac{\beta-1}{\beta+\gamma}.
\label{eq3}
\ee
Figure \ref{fig3} shows the thermal dependence of the averaged exponent $n$ for the studied compound estimated on the basis of the data in Fig. \ref{fig1}. In agreement with previous experimental data the general behaviour consists in  a smooth decrease of $n$ down to values close to the minimal value of 0.63 attained slightly above the transition temperature, and a subsequent increase toward $n=2$ being a consequence of the linear field dependence of magnetization where the Curie-Weiss law is valid. The value of $n$ at the transition temperature $T_\mathrm{c}=8.3$\,K is 0.64.
It was also demonstrated that the $\Delta S_\mathrm{M}(T)$ curves measured with different maximum applied fields can collapse into a single master curve when properly rescaled. A phenomenological recipe for doing this is to normalize all the $\Delta S_\mathrm{M}(T)$ curves with their respective peak entropy change and rescale the temperature axis as
\bd
\theta=\left\{\begin{array}{l}
-(T-T_\mathrm{c})/(T_{\mathrm{r}_1}-T_\mathrm{c})\quad T\leq T_\mathrm{c} \\
(T-T_\mathrm{c})/(T_{\mathrm{r}_2}-T_\mathrm{c})\quad T>T_\mathrm{c},
\end{array}\right.
\ed
where $T_{\mathrm{r}_1}$ and $T_{\mathrm{r}_2}$ are the temperatures of the two reference points that for the present study have been selected as those corresponding to 0.7$\Delta S_\mathrm{M}^\mathrm{peak}$. The selection of the factor has to be made in such a way that the curves to be overlapped have experimental values above that reference entropy change for temperatures below and above $T_\mathrm{c}$. Figure \ref{fig4} shows an attempt to form a master curve for the entropy change of the studied compound. One can see that the universal behaviour manifests itself only in a limited interval around the peak temperature. Farther below and above the peak the scaling behaviour apparently breaks. A similar lack of a complete overlapping of the rescaled magnetic entropy change curves below the transition temperature was reported for the amorphous alloy Er$_{0.15}$Dy$_{0.85}$Al$_2$ \cite{franco3} and it was argued to be due to a spin reorientation transition. However, the ac susceptibility signal reveals no anomaly below $T_\mathrm{c}$ down to 2\,K for the studied sample, see Fig. \ref{fig5}, which disproves some kind of a different magnetic phenomenon with different field dependence in this temperature regime. On the other hand, in terms of the reduced temperature $\varepsilon=(T-T_\mathrm{c})/T_\mathrm{c}$ the scaling region spans the interval $\varepsilon\in(-0.17,0.22)$, which is typical for systems with short-range interactions. Thus, the failure of the complete  scaling of $\Delta S_\mathrm{r}$ on both sides of the transition temperature may simply reflect the limited temperature regime where the critical behaviour holds.  Another reason for the lack of scaling on both sides of the transition temperature is the anisotropy of the Fe$^\mathrm{II}$ ions, but this could be only verified by the single crystal measurements.    

\begin{figure}[!ht]
\centering
\includegraphics[width=\textwidth]{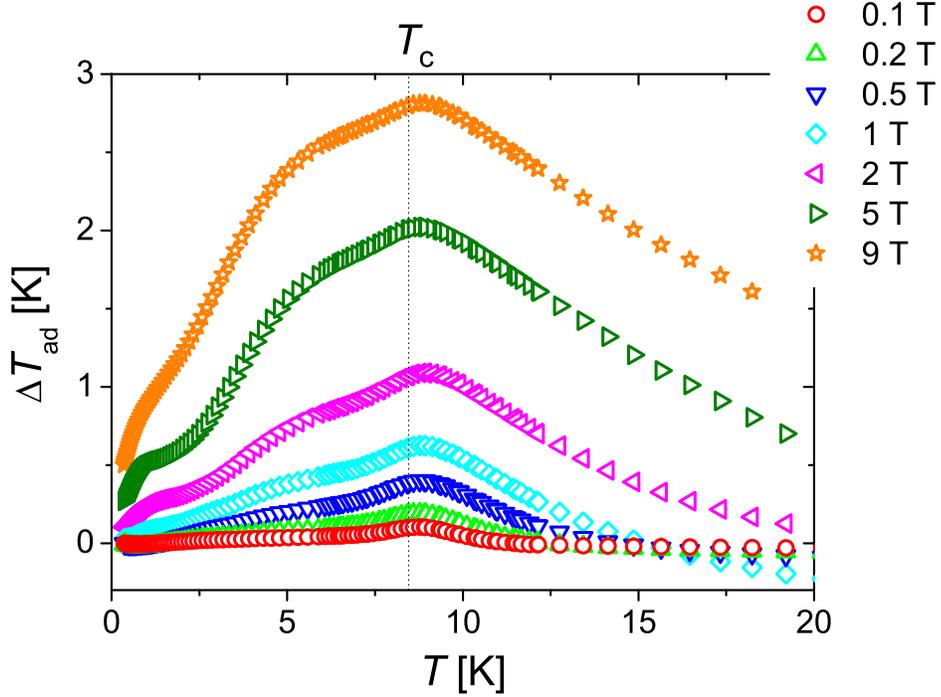}
\caption{Temperature dependence of the adiabatic temperature change $\Delta T_\mathrm{ad}$.}
\label{fig6}
\end{figure}

Finally, in Fig. \ref{fig6} the adiabatic temperature change is depicted as inferred from the formula given by Eq.(\ref{eq2}). Table \ref{tab1} lists the peak values of $\Delta T_\mathrm{ad}$ for $\mu_0\Delta H=$2, 5, and 9\,T. One can see that the peak positions do not depend on the applied field change and almost coincide with the transition temperature.  Like for the isothermal entropy change, the reverse MCE above 15\,K for low field values is also apparent. Below the transition temperature the $\Delta T_\mathrm{ad}$ vs. $T$  curves display visible irregularities with an inflection point at  about 4\,K and a sharp drop at the lowest temperatures. These features, especially apparent for higher field values, are signalling some field induced effect that may be related with the absence of scaling in this temperature regime. The values of $\Delta T_\mathrm{ad}$ are comparable to those reported for the isostructural compound \{[Ni$^\mathrm{II}$(pyrazole)$_4$]$_2$[Nb$^\mathrm{IV}$(CN)$_8$]$\cdot$4H$_2$O\}$_n$ (2.0\,K for $\mu_0\Delta H=$5\,T, and 2.9\,K for $\mu_0\Delta H=$9\,T) \cite{fitta2}, and larger than those found for Mn$_2$-pyridazine-[Nb(CN)$_8$] (1.5\,K for $\mu_0\Delta H=$5\,T) \cite{fitta1} and hexacyanochromate Prussian blue analogues (1.2\,K for $\mu_0\Delta H=$7\,T) \cite{manuel}.

\section{\label{section3}Conclusions}
Magnetocaloric effect for the unique molecular magnet \{[Fe$^\mathrm{II}$(pyrazole)$_4$]$_2$ [Nb$^\mathrm{IV}$(CN)$_8$]$\cdot$4H$_2$O\}$_n$ has been reported. We calculated the isothermal entropy change as well as the adiabatic temperature change using the results of the heat capacity measurements. The magnitudes of these quantities were found to be comparable with those reported for other representatives of molecule-based magnets. The knowledge of two critical exponents $\Delta$ and $n$ inferred from the field dependence of MCE is sufficient to determine other exponents, e.g. $\beta$ and $\gamma$. Using the relation $\Delta=\beta+\gamma$ and Eq.(\ref{eq3}) one easily finds $\beta=1-\Delta(1-n)\approx 0.35$, and $\gamma=\Delta(2-n)-1\approx 1.4$. These values are close to the theoretical estimates for the three-dimensional Heisenberg universality class \cite{campostrini}, namely $\beta=0.3689(3)$ and $\gamma=1.3960(9)$. This conclusion is consistent with the findings of the detailed study of this compound \cite{konieczny}, which shows that the presented analysis of the critical behaviour based on magnetothermal properties can yield reliable results. However, the issue of the lack of scaling further off the transition temperature remains still to be clarified. It seems clear that the single crystal measurements should shed some new light on that puzzle.











\end{document}